\documentclass{ar-1col-S2O}
\usepackage[numbers,sort&compress]{natbib}
\usepackage{url}
\usepackage{amsmath}
\usepackage{color}
\usepackage{hyperref}
\usepackage{gensymb}
\usepackage{pgfgantt}
\usepackage{pgfcalendar}

\jname{Annu. Rev. Nucl. Part. Sci.}
\jvol{76}
\jyear{2026}
\doi{10.1146/annurev-nucl-100324-
112126}

\begin{document}

\markboth{Martinez, Maruyama, and Sarsa}{Dark Matter Searches with NaI(Tl)}

\title{Dark Matter Searches with Sodium Iodide Detectors}

\author{Mar\'{\i}a Martinez$^1$, Reina H.~Maruyama$^2$, and Mar\'{\i}a Luisa~Sarsa$^1$ 
\affil{$^1$Centro de Astropart\'{\i}culas y F\'{\i}sica de Altas Energ\'{\i}as (CAPA), Universidad de Zaragoza, Pedro Cerbuna 12, 50009 Zaragoza, Spain: email: mariam@unizar.es, mlsarsa@unizar.es}
\affil{$^2$Wright Laboratory, Department of Physics, Yale University, New Haven, CT 06520, USA; email: reina.maruyama@yale.edu}
}

\begin{keywords}
sodium iodide scintillators, dark matter, direct detection, annual modulation, DAMA/LIBRA experiment
\end{keywords}

\begin{abstract}

The search for the identity of dark matter remains one of the most intriguing challenges in physics today, driving innovative advances in experiment and theory. Thallium-doped sodium iodide (NaI) detectors have played a pivotal role in this context. This review outlines the scientific landscape surrounding NaI detectors, emphasizing their role in the broader context of dark matter research. We begin with a brief description of characteristics that make these detectors well-suited for dark matter detection. We then provide a historical overview and explore the results from the DAMA/NaI and DAMA/LIBRA experiments, which have reported annual modulation signals consistent with dark matter interactions -- a signal that to date no other experiment has replicated. We discuss recent results from NaI-based experiments launched, in part, to resolve this tantalizing anomaly. We conclude with a summary of the status and prospects for current and future NaI-based dark matter searches.

\end{abstract}

\maketitle

\tableofcontents

\section{INTRODUCTION}

Sodium iodide (NaI) crystals, when doped with thallium [NaI(Tl)], are excellent particle detectors. Robert Hofstadter was the first to develop and patent NaI(Tl) detectors in 1948~\cite{NAS:2001aa}. 
The development of these detectors closely followed that of organic scintillators, which were first employed
to identify trace amounts of uranium during the Manhattan Project~\cite{curran-1949}, thereby improving the energy resolution and efficiency needed to detect $\gamma$-rays because of the higher-Z atom content.
Although NaI(Tl) is hygroscopic, when the crystals are encapsulated in aluminum or copper housing, they can be operated at room temperature and are field deployable. They are versatile and are used in many industrial, medical, nuclear, and high-energy physics applications, including well logging, medical imaging, and particle detection. They are also used in space missions, including as a calorimeter in EGRET (NASA's {\it Compton Gamma Ray Observatory} satellite and one of Hofstadter's projects;~\cite{Kanbach:1989cmd}) and as part of the Gamma-ray Burst Monitor in its successor {\it FERMI} satellite~\cite{Meegan:2009qu}. 
\begin{marginnote}
    \entry{NaI(Tl)}{thallium-doped sodium iodide}
    \entry{DM}{dark matter}
\end{marginnote}

The possibility of terrestrial-based direct detection of dark matter (DM) was proposed in 1985~\cite{Goodman:1984dc, Drukier:1986tm}. The results from the first search using germanium detectors were published in 1987~\cite{Ahlen:1987mn}, followed by several NaI(Tl)-based experiments in the 1990s~\cite{Beijing-Rome-SaclayBRS:1992ghk, Fushimi:1993nq,Sarsa:1996pa,Sarsa:1997hb}. 
In 1997, the DAMA/NaI collaboration~\cite{BERNABEI1998195} announced that it had observed an annual modulation in detection rates taken with 100\,kg of NaI(Tl) detectors, a finding consistent with what was expected from DM with the correct period and phase. 
In 2002, after running over seven annual cycles, the DAMA/NaI experiment confirmed the observation of an annual modulation and claimed the detection of DM~\cite{Bernabei:2003za}. DAMA/NaI was followed by Phase 1 of the DAMA/LIBRA ((Large sodium
Iodide Bulk for RAre processes) experiment using low background crystals (2003-2010)~\cite{Bernabei:2013xsa}.
Phase 2 of DAMA/LIBRA used an upgraded set of photomultiplier tubes (PMTs)~\cite{Bernabei:2011zzc}. The annual modulation signal persists throughout all phases of the experiment with 13.7$\sigma$ C.L.~-- a signal so large that it cannot be attributed to mere statistical fluctuation~\cite{Bernabei:2021kdo}.

During this time, many other direct detection DM  experiments were commissioned, and they have produced increasingly sensitive results. The experiments proposed and built from the 1990s to the 2010s focused on finding weakly interacting massive particles (WIMPs). WIMPs are a well-motivated DM candidate according to the so-called WIMP miracle, with masses ranging between 1\,GeV and  10\,TeV~\cite{Feng:2022rxt}. WIMPs interacting with atomic nuclei in the detector medium would produce a measurable energy release. The momentum transfer is most efficient when the mass of the nuclei is well-matched to the mass of the DM particle. 
A high detector mass and extremely low rates of background events are needed to carry out sensitive searches for DM. To minimize background rates, detectors are placed underground with a rock or water overburden and inside passive shielding, such as layers of neutron moderator, copper, and lead. Additional backgrounds may be tagged with an active veto, such as one made of plastic or liquid scintillators. 
\begin{marginnote}
    \entry{PMT}{photomultipler tube}
    \entry{Weakly interacting massive particle (WIMP)}{a DM candidate}
    \entry{CaWO$_4$}{calcium tungstate}
    \entry{LAr}{liquid argon}
    \entry{LXe}{liquid xenon}
    \entry{ER}{electron recoil}
    \entry{NR}{nuclear recoil}
\end{marginnote}

Targets such as argon, xenon, and germanium can be fashioned into highly sensitive particle detectors that excel in meeting these requirements. For instance, cryogenic bolometers [e.g., germanium, calcium tungstate (CaWO$_4$)] and liquid noble-gas detectors [e.g., liquid argon (LAr), liquid xenon (LXe)] enable the detection of two or more forms of energy released in the target: heat (phonons), light (photons), and charge (ionization electrons). By comparing the distribution of energy among these different channels, it is possible to distinguish between various types of particle interactions, such as electron recoil (ER), where energy is transferred to electrons, and nuclear recoil (NR), where energy is transferred to atomic nuclei.

Current-generation experiments (e.g., CRESST, SuperCDMS, DarkSide-50, DEAP-3600, Panda-X, XENON-nT, LZ) have reached sensitivities allowing them to rule out standard WIMPs as the cause of DAMA's observed annual modulation by orders of magnitude~\cite{CRESST:2019jnq, SuperCDMS:2017mbc, DarkSide-50:2022qzh, DEAP:2019yzn, PandaX:2024qfu, XENON:2024ijk, LZ:2024zvo, ParticleDataGroup:2024cfk}. The xenon experiments in particular are more than seven orders of magnitude more sensitive than DAMA, having reached the point where astrophysical neutrinos are starting to become the dominant source of background, yet they have found no evidence of DM~\cite{Billard:2021uyg, PandaX:2024muv, XENON:2025vwd}. 

New NaI(Tl) experiments have been deployed or are being planned. Their goals are twofold: {\it (a)} to exploit the improved performance of NaI detectors in light of the fact that NaI is a particularly attractive target material for DM searches beyond the standard WIMP scenario, where they are competitive, and {\it (b)} to test the validity of DAMA/LIBRA's assertion about DM detection using the same target medium and detection technique, removing dependencies on the DM and halo models. 
The ANAIS-112 (Annual Modulation with NaI Scintillators) and COSINE-100 experiments have made great strides toward the goal of testing DAMA. They recently published the results of their respective 6 year annual modulation searches~\cite{Amare:2025dfq, COSINE-100:2024nfa}. Neither experiment found an annual modulation, and when combined, their data exclude DAMA/LIBRA's signal at 4.7$\sigma$ for 1--6\,keV and 3.5$\sigma$ for 2--6\,keV~\cite{ANAIS-112:2025fne}. 

The lack of signal from such a wide variety of sensitive direct detection DM experiments has spurred much debate and raised a new set of questions.
DM models favoring sodium or iodine over xenon or germanium have been proposed; however, they are now severely constrained, as the experimental limits have become more stringent~\cite{Billard:2021uyg, PandaX:2024muv, XENON:2025vwd}. Additionally, the lack of modulation found by ANAIS-112 and COSINE-100 provides a strong model-independent argument against a DM  interpretation of the DAMA/LIBRA signal. As various experimental efforts approach WIMP sensitivities that are constrained by the neutrino floor, the community's attention has shifted to other DM candidates such as axions, axion-like particles, dark photons, and other low-mass particles~\cite{Graham:2015ouw, Choi:2020rgn, Graham:2021ggy, Hui:2021tkt, Berlin:2024pzi}. Many WIMP detectors can be adapted to search for low-mass DM by lowering the detector energy threshold and either employing lighter nuclei or searching for collisions with electrons~\cite{Zurek:2024qfm, Essig:2022dfa}.

In this article, we review the history and current status of DM searches with NaI detectors. In Section~\ref{sec:current} we summarize the current experimental status of NaI-based DM searches. In Section~\ref{sec:interpret} we first provide a brief overview of the model-dependent exclusions of DAMA/LIBRA's annual modulation signal and of the various interpretations proposed to date. We then describe the status of the model-independent rejection of the DAMA/LIBRA signal and examine in detail the systematic effects that influence comparisons among NaI(Tl) experiments. In Section~\ref{sec:projects} we present the NaI projects currently being commissioned, in advanced development, or in the research and development (R\&D) phase. We conclude in Section~\ref{sec:Conclusions} with a summary of the status of NaI searches and an outlook on possible paths toward a definitive resolution of the anomaly of the DAMA/LIBRA signal.

\section{SODIUM IODIDE DETECTORS AND DETECTION OF DARK MATTER}
\label{sec:current}

\subsection{Direct Detection Dark Matter Searches}

Various observational and experimental approaches have been employed to understand DM beyond its gravitational effects. These include the use of particle colliders, as well as indirect and direct detection methods. In particle colliders, high-energy collisions might produce DM particles or reveal signs of new physics that could explain DM. Indirect detection involves searching for the decay or annihilation products of DM in different astrophysical environments, which would result in excesses of known particles, such as $\gamma$-rays, neutrinos, or charged particles, above other astrophysical backgrounds.
In contrast, direct detection, which is the main focus of the review, involves using particle detectors on Earth to look for recoils of nuclei or electrons within the detector medium caused by interactions with DM particles pervading the Milky Way halo. In the case of searches for axions (which are also a good DM candidate, but a topic for another review), detection relies on specific interaction channels such as axion-photon coupling (resulting in axion-photon conversion mediated by virtual photons) or axion-electron coupling~\cite{Graham:2015ouw, Choi:2020rgn, Berlin:2024pzi}. These strategies are complementary, and together they are contributing to the solution of the most intriguing puzzles in cosmology and particle physics: the nature of DM.

In direct DM searches, the objective is to detect the energy transferred directly from DM particles to nuclei or electrons in the target material, which results in NR or ER, respectively. The amount of energy transferred and the anticipated interaction rate depend strongly on the properties of the DM particles such as their mass and couplings to quarks and/or leptons, as well as on the density and velocity distribution of DM particles at the Solar system position. In this mode of detection, DM particles with masses between 1\,GeV and 10\,TeV provide the most efficient energy transfer. Uncertainties in the galactic halo model are important but are somewhat constrained. Observations and N-body simulations provide a good description of the DM density and velocity distribution in the galactic reference frame. However, substructures within the halo and potential directional fluxes might exist and influence the expected interaction rates. 

The properties of DM particles significantly affect direct detection searches, particularly the nature of their coupling with Standard Model particles. In most WIMP and WIMP-like models, these particles are expected to produce NR through either spin-independent (SI) or spin-dependent (SD) coupling to nucleons. However, up to 15 effective operators could be present in the Lagrangian, including several that depend on the WIMP's velocity~\cite{Fitzpatrick:2012ix}. Interpreting experimental results in terms of all possible couplings presents a considerable challenge. 
Additionally, NR can induce atomic ionization or excitation through the Migdal effect~\cite{Ibe:2017yqa}, leading to the emission of ionized electrons and rendering otherwise subthreshold NRs observable in the ER channel. This phenomenon was recently confirmed experimentally~\cite{Yi:2026fmf}, and the measured rates were consistent with theoretical expectations. As more precise measurements become available, their quantitative impact on the sensitivities of present and future DM direct searches should be further assessed.
Moreover, other DM candidates beyond WIMPs might directly produce ER by coupling to electrons.

\begin{marginnote}
    \entry{Spin-independent (SI) coupling}{DM coupling to nucleons, which adds coherently over the nucleus, although loss of coherence can reduce the signals if momentum transfer is high}
    \entry{Spin-dependent (SD) coupling}{DM coupling to the spin of the nucleons within the nucleus}
\end{marginnote} 

Depending on the detection technique applied, the conversion of the energy released to visible energy can differ for NR and ER, requiring different energy calibration techniques. In situ detector calibration is usually carried out in ER because full-energy $\gamma$  peaks are readily accessible, are monoenergetic, and can easily penetrate deep into detectors. NR calibration is more difficult and typically requires the use of a neutron source. Experiments use a conversion factor, the quenching factor (QF), to convert between the ER and NR energy scales. The QF is notoriously difficult to measure and makes it more challenging to compare results from experimental searches using different target nuclei and detection techniques. Such comparisons are inherently model dependent because they require that we choose a specific DM particle candidate and galactic halo model.

\begin{marginnote}
    \entry{Quenching Factor (QF)}{scaling factor used to account for the difference in the energy calibration between ER and NR events; scintillation QF measures the relative light output of ER versus~NR releasing the same energy}
\end{marginnote} 

Conventional direct detection DM experiments look for event count rates that exceed all known backgrounds and, therefore, could be attributable to DM particles scattering off of atomic nuclei. These experiments optimize signal detection by maximizing the number of atomic nuclei (detector mass) and minimizing backgrounds. They achieve this goal by operating underground, using radiopure materials in the detector building, surrounding the detector with passive shielding and an active veto against external backgrounds, and discriminating NR from ER when possible.

The DM count rate is expected to exhibit an annual modulation due to the Earth's motion around the Sun (for a review, see~\cite{FreeseColloquium}). This motion leads to a modulation in the relative velocity between the Earth and the DM halo, which in turn changes the flux of particles arriving at the detector and, therefore, the count rate.
By the beginning of June (December), the Earth is moving in the same (opposite) direction as the Sun and the relative velocity is highest (lowest), resulting in a peak (trough) in the count rate for the higher energy depositions; the effect is the opposite at the lowest energies. It is possible for the signal to be flat when there is an unfortunate cancellation between signals from two different target masses, one large and one small, that produces the opposite modulation.

Diurnal modulation is also expected as a result of the Earth's rotation on its axis, but this effect is small compared with the annual modulation, since the rotational velocity of the Earth is 0.5\,km\,s$^{-1}$ near the equator as opposed to the Earth's orbital velocity of 30\,km\,s$^{-1}$s. However, strong forward-backward asymmetry in the direction of the DM particle wind is expected. Detection techniques that are sensitive to the direction of a low-energy NR are currently under development~\cite{Miuchi:2023act}. 
These modulations provide an additional means of verifying that an excess in the count rate corresponds to a DM signal, as they are characteristic of DM interactions and are not typically associated with most known background sources. Observing these periodic variations will be crucial for confirming the origin of a potential signal as DM by differentiating it from background noise.

Reaching the highest sensitivity to light-mass DM particles in the GeV mass range also requires reducing the detectors' energy threshold. The challenges are different from those faced by the high-mass DM experiments, and leading light-mass DM experiments still have small detector masses~\cite{Billard:2021uyg}. Detector technology that does not rely on momentum transfer to nuclei is needed~\cite{Essig:2022dfa} in order to detect DM in the 1\,meV--1\,GeV range.

\subsection{Sodium Iodide Detectors}
\label{sec:NaIDetectors}
NaI is a particularly attractive target material for DM direct detection searches. High-quality, high-purity, and high-mass NaI(Tl) detectors can be grown at relatively low cost. Although NaI(Tl) is hygroscopic, when the crystals are encapsulated, they can be operated at room temperature and are field deployable. Detectors are easy to operate and stable, and they can run with a relatively short dead time and high duty cycle. 

NaI(Tl) detectors have excellent light output and can produce up to 40,000\,photons\,MeV$^{-1}$. Their emission spectrum is well-matched to the PMT sensitivity curve, leading to a typical energy resolution of 6\% at full width at half-maximum (FWHM) at 662\,keV and enabling an energy threshold below 1\,keV. 
The ANAIS-112 and COSINE-100 collaborations have achieved light collection at the level of 15 and 22\,npe\,keV$^{-1}$ (where npe refers to the number of photoelectrons), respectively, with 12.5\,kg mass crystals~\cite{Olivan:2017akd} and up to 22~npe\,keV$^{-1}$ in smaller crystals~\cite{Choi:2020qcj}.
NaI(Tl) light output exhibits several decay times, with a primary time constant of $\sim$250\,ns at room temperature as well as longer-lived phosphorescence~\cite{Cuesta:2013vpa}. The light output per keV of energy deposition shows nonproportional behavior at the level of a few per cent, and it differs between ER and NR by a QF. 

\begin{marginnote}
    \entry{Number of photoelectrons (npe)}{number of photons emitted by the scintillator that reach the PMT photocathode and generate a signal}
    \entry{Pulse-shape discrimination (PSD) or pulse-shape analysis (PSA)}{PSD or PSA techniques enable discrimination of, for instance, NR from ER according to their different pulse shapes}
\end{marginnote}

The NaI(Tl) detectors used in DM searches are typically coupled to two or more PMTs that trigger in coincidence to minimize single photoelectron noise events (dark counts) and increase light collection. Low-background materials such as polytetrafluoroethylene (PTFE), copper, and fused-silica are used as a light reflector, for encapsulation, and as a light-coupling window, respectively.

Regarding sensitivity in DM detection, the presence of both a heavy and a light nucleus provides favorable kinematic matching over a wide range of WIMP masses. In addition, 100\% of the detector is made of isotopes with nonzero nuclear spin ($^{127}\mathrm{I}$ and $^{23}\mathrm{Na}$) and, therefore, is sensitive to SD-interacting WIMPs. Pulse-shape discrimination (PSD) between ER and NR is possible, although it is challenging below 4\,keV and requires more R\&D before it can be used reliably in DM searches. QFs for sodium and iodine recoils in NaI(Tl) are required in order to calibrate the NR energy scale, enabling comparisons between results from different detectors and techniques. At present, we do not know whether the QF is an intrinsic characteristic of the material or whether it depends on the particular \mbox{crystal properties (see Section~\ref{sec:QF})}.

\subsection{Direct Detection Pioneers: Early Sodium Iodide Experiments}
\label{sec:pioneers}

NaI detectors were adapted as DM detectors early on and laid the foundation for future large-scale WIMP DM experiments. In particular, in the early 1990s the BPRS (Beijing-Paris-Rome-Saclay), DM32, and ELEGANTS-V (Electron Gamm-Ray Neutrino Spectrometer V) Collaborations assembled a few tens of kilograms of low-background NaI(Tl) detectors for use as DM detectors at their respective underground laboratories. These experiments were among the largest and most sensitive at the time and directly competed with the leading germanium-based DM experiments. 

\subsubsection{BPRS Collaboration}
\label{sec:BPRS}
In the early 1990s, the BPRS Collaboration conducted pioneering direct searches for DM using NaI(Tl) scintillators. The collaboration developed an ambitious experimental program at three underground laboratories: Gran Sasso National Laboratory (LNGS), Fréjus, and Mentogou. After preliminary measurements, BPRS achieved a low-energy event rate with sensitivities to SD-interacting WIMPs that were comparable to those of the leading germanium experiments~\cite{Beijing-Rome-SaclayBRS:1992ghk}. This collaboration was the seed of the subsequent DAMA/NaI experiment.

\subsubsection{DM32}
\label{sec:DM32}
Around the same time, members of the Nuclear and Astroparticle Physics Group at the University of Zaragoza in Spain assembled and carried out the {DM32 Experiment} at Canfranc Underground Laboratory (LSC). The experiment consisted of three 10.7\,kg NaI(Tl) modules with a total mass of 32.1\,kg. The detectors were originally deployed as an active veto for a germanium-based neutrinoless double-$\beta$-decay experiment~\cite{Morales:1991mh}. With 4,613.6\,{kg}$\cdot${day} of data taken between May 1993 and December 1994, DM32 set competitive limits on SI and SD coupling between DM particles and target nuclei, by looking for both an excess in the total number of events and an annual modulation in the event rate. The latter allowed for an improvement of approximately two orders of magnitude in the parameter space ruled out by the experiment~\cite{Sarsa:1996pa,Sarsa:1997hb}.

\subsubsection{NaI from ELEGANTS-V}
\label{sec:ELEGANTS}
Of the 20 NaI moduless used for the ELEGANTS-V experiment, 9 were assembled in 1992 and deployed at Kamioka Underground Laboratory. The central detector with a mass of 36.5\,kg, was used for the DM search, while the others were deployed as active shielding~\cite{Ejiri:1991ct, Fushimi:1993nq}. The noise threshold was 4\,keV, and the background rate at 5\,keV was around 5.1\,cpd\,kg$^{-1}$\,keV$[^{-1}$ (where cpd stands for counts per day). For the first time, the scintillation QFs were determined by exposing the crystal to a $^{252}$Cf source and fitting the elastic scattering spectral shape. This procedure was followed by the DAMA/NaI and ANAIS-112 experiments (see Section~\ref{sec:QF}).

\begin{marginnote}
    \entry{cpd\,kg$^{-1}$\,keV$[^{-1}$}{counts per day of measurement per kilogram of detector mass and per keV of energy deposited}
    \entry{Meters of water equivalent (m.w.e)}{overburden equivalence in meters of water}
\end{marginnote} 

\subsection{DAMA/NaI, DAMA/LIBRA, and the Claim of Detection of Dark Matter}
\label{sec:claim}
The DAMA Collaboration carried out a collection of experiments to investigate the identity of DM and to search for other rare processes, such as neutrinoless double-$\beta$ decay. While other experiments, such as DAMA/LXe and DAMA/Ge, have produced physics results, DAMA/NaI and its successor, DAMA/LIBRA, are of interest for this review.
\subsubsection{DAMA/NaI}
\label{sec:DAMA/NaI}
Some of the members of BPRS went on to form a new collaboration, DAMA/NaI which ran from 1995 to 2002 at LNGS. LNGS is located approximately 100\,km east of Rome, Italy, under 1,400\,m of rock overburden, equivalent to an average depth of 3,800\,meters of water equivalent (m.w.e.). As such, it provides substantial shielding from cosmic rays, reducing the muon flux by a factor of approximately $10^6$. 

DAMA/NaI was made of nine 9.7\,kg NaI(Tl) crystals, each coupled to two PMTs through 10\,cm of light guides made of quartz. The detectors were wrapped with PTFE light reflectors and encapsulated in copper. The detectors were inside a sealed copper box flushed with nitrogen and surrounded by copper, lead, polyethylene/paraffin, cadmium foil, and a 1\,m concrete neutron moderator as shielding for $\gamma$ rays and neutrons. 

The collaboration first reported a hint of an annual modulation in 1997~\cite{BERNABEI1998195} and for the full data set in 2003, over seven annual cycles corresponding to a total exposure of 107,731\,kg$\cdot$day~\cite{Bernabei:2003za}.
It observed an annual modulation with a period and phase consistent with DM at 6.3$\sigma$ CL. The analysis used the residuals in the 2--6\,keV energy region from all nine modules, defined as the time-dependent deviations of the event rate from its mean value, and fitted them to a cosine function:
\begin{equation}\label{eq:modulation}
    S(t) = S_m cos[\omega (t-t_0)],
\end{equation}
where \(S_m\) is the modulation amplitude, \(\omega = 2\pi / T\) is the angular frequency corresponding to the expected annual period \(T = 1~\mathrm{year}\), and \(t_0\) is the phase of the modulation.  

DAMA/NaI reported a modulation with an amplitude of $S_m = 20\pm3.2$\,cpd\,ton$^{-1}$\,keV$^{-1}$, a period of $T=1.00\pm0.01$\,year, and phase of $t_0=140\pm22$\,days, with all the parameters set floating in the fit. If the period and phase were fixed in the fit to $T=1$~year and $t_0=$~June 2$^\text{nd}$ (corresponding to $t_0=152$~days), then the modulation amplitude of the best-fit in the same region was $19.2\pm3.1$\,cpd\,ton$^{-1}$\,keV$^{-1}$ for single-hit events and was absent for multiple-hit events, as expected for WIMP-induced interactions. In contrast, the same analysis applied in the 6--14\,keV region resulted in a modulation compatible with zero, with a best-fit amplitude of $-0.9\pm1.9$\,cpd\,ton$^{-1}$\,keV$^{-1}$. The annual modulation signal was analyzed within various astrophysical and particle physics frameworks, including SI, SD, mixed, and inelastic couplings, leading to an allowed region in the WIMP parameter space that is compatible with many viable DM candidates~\cite{Bernabei:2003za}.

\begin{marginnote}
    \entry{Single-hit events}{events with only one NaI(Tl) module triggering}
    \entry{Multiple-hit events}{events with more than one trigger. In DAMA/NaI, there was no active veto; therefore, multiple-hit events correspond to at least two NaI(Tl) modules triggering}
\end{marginnote} 

The DAMA/NaI collaboration examined potential contributions from radon activity, temperature  fluctuations, noise or electronic instabilities, calibration and efficiency drifts, and background variations due to neutrons or muons in the LNGS~\cite{Bernabei:2000ew}. None of these sources exhibited a modulation that could mimic the observed signal; the estimated impact on the count rate from these sources was at least two orders of magnitude smaller than the measured modulation amplitude. 

\subsubsection{DAMA/LIBRA} 
\label{sec:DAMA/LIBRA}

Following DAMA/NaI, the DAMA/LIBRA experiment was installed at LNGS. The experiment consisted of 25~highly radiopure NaI(Tl) crystal detectors, each with a mass of 9.7\,kg in a 5$\times$5 configuration, for a total mass of $\sim 250$\,kg. Detectors were produced following an R\&D program developed in collaboration with Crismatec/Saint-Gobain, which yielded improved radiopurification techniques for the selected NaI and TlI (thallium iodide) powders. Additional protocols were implemented to further enhance the overall radiopurity~\cite{DAMA:2008bis}.

DAMA/LIBRA took data in two phases from 2003 to 2024. Phase~1 ran from 2003 to 2010 and had an energy threshold of 2\,keV. Phase~2 ran from 2010 to 2024. For Phase~2, the PMTs were replaced with high-quantum-efficiency PMTs (Hamamatsu R6233MOD)~\cite{Bernabei:2012zzb}. These PMTs enabled better overall light collection and a lower background rate, resulting in a lower energy threshold of 1\,keV. 

The results from DAMA/LIBRA-Phase~1 were reported for four~\cite{DAMA:2008jlt}, six~\cite{DAMA:2010gpn} and seven~\cite{Bernabei:2013xsa} annual cycles. 
The analysis strategy involved building the residual rate of the single-hit scintillation events in the 2--6\,keV energy region for the full exposure. The average measured rate was subtracted for each data set, which lasted approximately 1 year.  
The remaining residual rate was fitted to a cosine dependence (Equation~\ref{eq:modulation}) with the modulation amplitude $S_m$ floating, while the period (T = 2$\pi/\omega$) and the phase (t$_0$) were either floating or fixed at 1~year and June 2. When the period and phase were floated, the experiment obtained values compatible with those expected for DM within the standard halo model. In the fit with a fixed period and phase, it obtained modulation amplitudes of $S_m = 9.6\pm1.3$\,cpd\,ton$^{-1}$\,keV$^{-1}$ at 7.4$\sigma$ CL~in the 2--6\,keV energy region with $\chi^2/dof=29.3/49$~\cite{Bernabei:2013xsa}. 
This analysis procedure is known to be affected by systematics if time-dependent backgrounds are present~\cite{Messina:2020pnt,Buttazzo:2020bto,COSINE-100:2022dvc}. The DAMA/LIBRA Collaboration argued that data taking started in the fall when the rate from DM is expected to be low. Therefore, the proposed systematic could not explain their result, which woud require a background in the region of interest (ROI) that increases steadily with time.  

DAMA/LIBRA-Phase~2 presented results after seven additional annual cycles~\cite{Bernabei:2018jrt, Bernabei:2020mon}. The reported modulation amplitudes for fixed period and phase were
$S_{m} = 10.2 \pm 0.8~\mathrm{cpd\,ton^{-1}\,keV^{-1}}$ with a significance of
$12.8\sigma$ in the 2--6~keV region (using the whole exposure, including DAMA/NaI), and
$S_{m} = 10.5 \pm 1.1~\mathrm{cpd\,ton^{-1}\,keV^{-1}}$ with a significance of
$9.5\sigma$ in the 1--6~keV region (using DAMA/LIBRA--Phase~2 only)~\cite{Bernabei:2020mon}. 

\ganttset{
    dama/.style={bar/.append style={fill=blue!35, draw=blue}},
    naiad/.style={bar/.append style={fill=magenta!35, draw=magenta!70!black}},
    dmice/.style={bar/.append style={fill=cyan!35, draw=cyan!70!black}},
    cosine/.style={bar/.append style={fill=green!35, draw=green!50!black}},
    anais/.style={bar/.append style={fill=orange!40, draw=orange!80!black}}
}

\begin{figure}[thb]
\centering



\begin{ganttchart}[
    x unit=0.3cm,
    y unit title=0.6cm,     
    y unit chart=0.5cm,
    vgrid={*4{draw=gray!30}, draw=black},
    title label font=\scriptsize,
    bar top shift=0.4,
]{1993}{2027}


\gantttitlelist[
    title height=1,   
    title label font=\tiny,
    title label node/.append style={
        rotate=75,
        anchor=west,
        inner sep=0pt,
        xshift=-5pt 
    },
    title/.append style={draw=none},
]{1993,...,2027}{1} \\

\ganttbar[anais]{DM32}{1993}{1994} \\
\ganttbar[dama]{DAMA/NaI}{1995}{2002} \\
\ganttbar[naiad]{NAIAD}{2000}{2003} \\
\ganttbar[dmice]{DM-Ice17}{2011}{2015} \\
\ganttbar[dama]{DAMA/LIBRA}{2003}{2024} \\
\ganttbar[cosine]{COSINE-100}{2016}{2023} \\
\ganttbar[anais]{ANAIS-112}{2017}{2025}

\end{ganttchart}
\caption{Timeline of NaI(Tl) dark matter experiments from 2013 to 2027, showing the data taking-periods.}
\label{fig:nai_timeline}

\end{figure}

\subsection{NaI(Tl) Experiments Testing DAMA/LIBRA's Claim of Dark Matter Detection}
\label{sec:testing}

\begin{table}[]
    \caption{Effective mass, exposure, energy threshold, and search method of pioneering and current leading NaI dark matter experiments}
    \label{tab:experiments}
    \begin{tabular}{c|c|c|c|c}
    \centering
      Experiment   &  Mass & Exposure  & Threshold & Signal\\
            & (kg) & ($\text{kg}\cdot\text{yr}$) & (keV) & (AM/TR/PSA)\\
    \hline
  DM32~\cite{Sarsa:1997hb}   & 32.1  & 12.64  & 8.0 & AM/TR\\
  DAMA/NaI~\cite{Bernabei:1996vj,DAMA:2000mdu} & 97 & 295.15  & 2.0 & AM/TR wPSA\\
  NAIAD~\cite{UKDarkMatter:2005xns}  & 55 & 44.9 & 4.0 & TR wPSA\\
  DM-Ice17~\cite{DM-Ice:2016snk} & 17 & 60.8 & 4.0 & AM/TR\\
  DAMA/LIBRA (full dataset)~\cite{Bernabei:2020mon} & 242.5 & 2462.09 & 2.0 & AM\\
  DAMA/LIBRA-Phase 2~\cite{Bernabei:2020mon} & 242.5 & 1126.40 & 1.0 & AM\\
  COSINE-100~\cite{COSINE-100:2024nfa} & 61.3 & 358 & 1.0 & AM/TR\\
  ANAIS-112\cite{Amare:2025dfq} & 112.5 & 625.75 & 1.0 & AM\\
  \hline
    \end{tabular}
    \vspace{0.2cm}
    \begin{quote}
    \small{Abbreviations: AM, annual modulation search; TR, total count rate, wPSA with pulse-shape analysis} 
    \end{quote}
    
\end{table}

\begin{figure}
    \centering
    \includegraphics[width=0.75\textwidth]{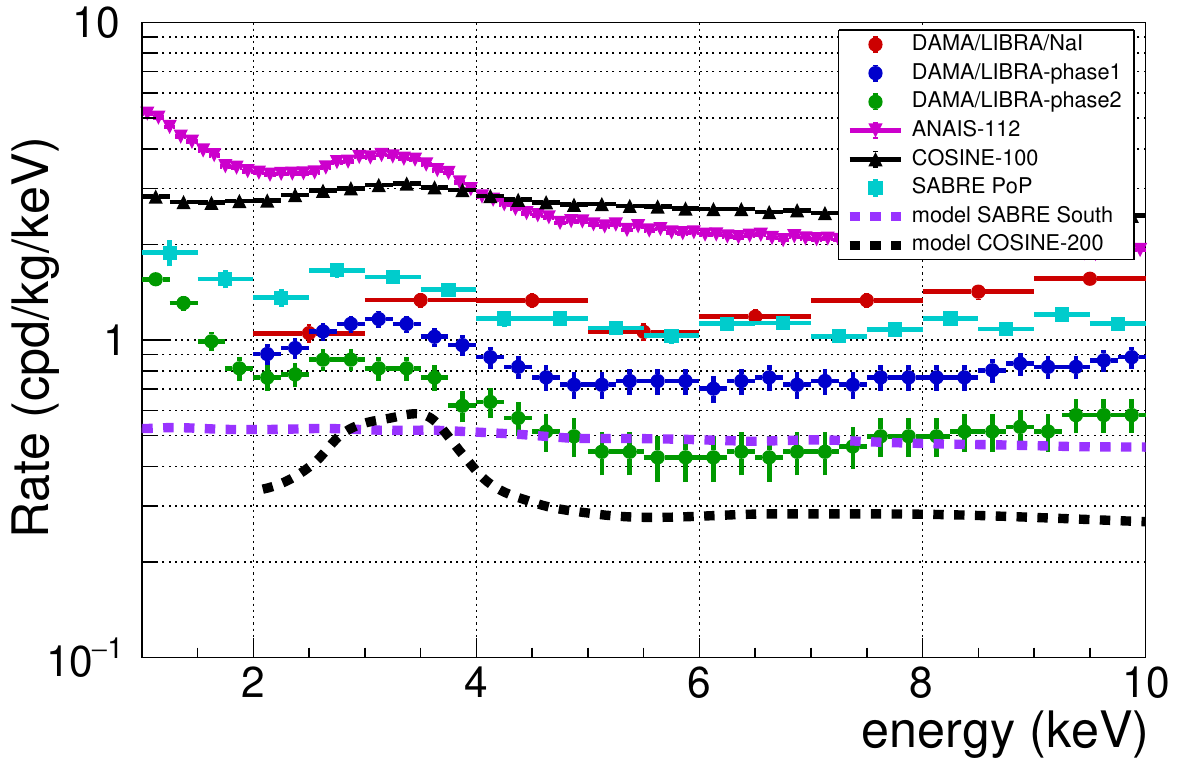}
    \caption{Comparison of the total single-hit energy spectra measured after six years of data by ANAIS--112 ({\it magenta triangles})~\cite{Amare:2025dfq}, COSINE--100 ({\it black triangles})~\cite{COSINE-100:2021xqn}, DAMA/LIBRA (Phase 1, {\it blue dots}; Phase 2, {\it green dots})~\cite{Bernabei:2020mon} , DAMA/NaI ({\it red dots})~\cite{BERNABEI1998195},  and SABRE proof of principle ({\it cyan squares})~\cite{Calaprice:2022vte}, along with the background models estimates from SABRE-South ({\it violet dotted line})~\cite{SABRESouth:2024bpv} and COSINE-200 ({\it black dotted line})~\cite{Lee:2023jbe}.}
     \label{fig:bkg}
\end{figure}

To test the DAMA/LIBRA claim, several groups have mounted or are mounting experiments using the same target material. Table~\ref{tab:experiments} and Figure~\ref{fig:nai_timeline} summarize the relevant experimental features and the timelines of the pioneering and current NaI DM experiments. If the DAMA/LIBRA signal is produced by DM, then the same modulation in terms of amplitude, period, and phase should be observed anywhere on Earth. If the signal depends on a seasonal variation that has not yet been determined, then different modulations or even an absence of modulation should be observed at different locations. Replicating the DAMA/LIBRA experiment has proved difficult to achieve. The NaI radiopurity reported by the DAMA/LIBRA Collaboration remains among the best to date (Figure~\ref{fig:bkg}). However, and partly because DAMA/LIBRA's signal is so large, the ANAIS-112 and COSINE-100 Collaborations have constructed detectors able to test the DAMA/LIBRA claim directly, as described below.

\subsubsection{NAIAD} 
\label{sec:NAIAD}
The NAIAD (NaI Advanced Detector) experiment was mounted by the UK Dark Matter Collaboration at the Boulby Mine underground site, at a vertical depth of 2,805~m.w.e. With 6~months of data from a 6\,kg NaI crystal, NAIAD obtained a limit for SD-interacting WIMPs that was a factor of 50 better than the state of the art, obtained with a germanium detector~\cite{Smith:1996fu}. NAIAD used PSD to separate NR from ER events in the 4--25\,keV range.

From 1997 to 2000, the discovery of fast anomalous scintillation events in data from several encapsulated crystals --faster than ER and NR-- prevented the experiment from reaching the intended sensitivity. These events were later found to be caused by surface contamination from radon decay.

From 2000 to 2003, an array of several NaI(Tl) crystals with a total mass of $\sim$55\,kg was installed. NAIAD analyzed $44.9~\text{kg}\cdot\text{yr}$ of data corresponding to two encapsulated and four unencapsulated NaI(Tl) crystals. The lack of an observed signal from NR set upper limits on the WIMP-nucleon SI and WIMP-proton SD cross sections, which were the most stringent at the time~\cite{UKDarkMatter:2005xns} but also compatible with the parameter space region that could explain the DAMA/NaI annual modulation result. Some of these UK Dark Matter Collaboration crystals were later used in the DM-Ice experiment. 

\subsubsection{DM-Ice17}
\label{sec:DM-ice}

In December 2010, during the last IceCube construction season, DM-Ice17 was deployed 2,500\,m deep in the Antarctic ice at the South Pole; it began taking data in June 2011. DM-Ice17 consisted of two modules --each equipped with an 8.5\,kg NaI(Tl) crystal, two PMTs, and two IceCube-style electronics boards to supply power to and read out the signals from the PMTs-- with a total of 17\,kg of NaI~\cite{DM-Ice17:2014xyz}. Each module was housed in a stainless steel pressure vessel to withstand up to 10,000\,psi from the water above and the pressure that can build up in the surrounding water as the ice freezes. The NaI crystals were donated by the NAIAD Collaboration~\cite{UKDarkMatter:2005xns}. One detector was deployed at the bottom of an IceCube string near the center and the other 500\,m away, toward the edge of IceCube. Although the total mass was low and the background too high to test the DAMA/LIBRA claim, DM-Ice17 demonstrated the feasibility of deploying low-background scintillation detectors in the ice at the South Pole~\cite{Cherwinka:2011ij}, set modest limits on DM particles, and investigated annual modulations from muons at the South Pole~\cite{Cherwinka:2011ij, DM-Ice17:2014xyz, DM-Ice:2015aij, DM-Ice:2016snk}. 

\subsubsection{KIMS}
\label{sec:KIMS}
Although technically not a NaI experiment, the KIMS (Korean Invisible Mass Search) experiment used $\sim$100\,kg of CsI(Tl) crystals at the Yangyang Underground Laboratory (Y2L) to test the hypothesis that DAMA/LIBRA's signal is dominated by DM's interactions with iodine. In 2007, with an exposure of 3,409\,kg$\cdot$day, KIMS ruled out WIMPs or WIMP-like particles with mass greater than $20$\,GeV as the cause of DAMA/LIBRA's signal~\cite{KIMS:2007wwj}. A stronger exclusion was set in 2012 with data collected between September 2009 and August 2010 corresponding to 24,524.3\,kg$\cdot$day of exposure~\cite{Kim:2012rza}. This test of the DAMA/LIBRA's claim is affected by the uncertainties in the QF (we discuss this topic further in Section~\ref{sec:QF}).

\subsubsection{COSINE-100}
\label{sec:COSINE100}
The COSINE-100 experiment was designed to directly test DAMA/LIBRA's claim of DM detection. COSINE-100 ran from October 2016 to March 2023 and was hosted at Y2L, which provides a granite overburden of minimum of 700\,m. The experiment consisted of eight NaI(Tl) crystals with a total mass of 106\,kg, however three crystals were used for tagging but were removed from the physics analyses because two of them had poor light collection and one had unexpectedly large noise from its PMTs. The total mass used for the physics analyses was 61.4\,kg. The crystals were produced by Alpha Spectra Inc. as part of an R\&D campaign by the KIMS and DM-Ice collaborations. The detectors were immersed in $\sim$2,200~L of liquid scintillator that acted as an active veto and were housed in an acrylic box surrounded by 3\,cm of copper, 20\,cm of lead, and 3\,cm of plastic scintillator as a muon veto. COSINE-100 had a background rate of 3--4\,cpd\,kg$^{-1}$\,keV$^{-1}$ in 2--6\,keV~\cite{Adhikari:2017esn} (Figure~\ref{fig:bkg}). 

In 2018, COSINE-100 published the results of its first search for SI-interacting DM, which focused on finding events in excess of the background by using only the first 59.5\,days of data and an analysis threshold of 2~keV~\cite{Adhikari:2018ljm}. No evidence for DM was found, and the data strongly disfavored SI DM as the origin of DAMA/LIBRA's annual modulation signal. To enable this analysis, COSINE-100 developed a detailed background model using Geant4-based simulations~\cite{COSINE-100:2018tfl}. In 2021, COSINE-100 published a pair of papers~\cite{COSINE-100:2021mrq, COSINE-100:2021xqn} that included 1.7\,years of data, leading to improvements in the understanding of the backgrounds and a more robust exclusion of DAMA/LIBRA signal that included even the most pessimistic QF (see also Section~\ref{sec:QF}). The models included time-dependence, which is crucial for the annual modulation searches. Background modeling has now been extended to 0.7--4,000\,keV~\cite{COSINE-100:2024ola}; 3 years of data provided an even stronger constraint on WIMP DM with the total rate analysis~\cite{COSINE-100:2025kbw}. 

In addition, COSINE-100 carried out searches for annual modulation. The first search was published in 2019 with 1.7\,years of data, 97.7\,$\text{kg}\cdot\text{yr}$ of exposure, and a 2.7\,cpd\,kg$^{-1}$\,keV$^{-1}$ background rate on average at \mbox{2--6\,keV}~\cite{COSINE-100:2019lgn}. The second data set was published in 2021 with 2.82~years of data and 173\,$\text{kg}\cdot\text{yr}$ of exposure, and it included improved event selection with a lower energy threshold of 1\,keV and an improved time-dependent background model in the 1--6 and 2--6\,keV energy intervals~\cite{COSINE-100:2021zqh}. The third and final data set was published in 2025 with 6.4\,years of data and 358\,$\text{kg}\cdot\text{yr}$ of exposure~\cite{COSINE-100:2024nfa}. COSINE-100 found no evidence of an annual modulation signal, challenging the DAMA/LIBRA result with $>$3$\sigma$ CL. The COSINE-100 experiment concluded in March 2023 after 6.4\,years of stable operation.

\subsubsection{ANAIS-112}
\label{sec:ANAIS112}
The ANAIS-112 experiment is located in Hall B of LSC in Spain, under a rock overburden of 2,450\,m.w.e. The experiment started taking physics data on August 3, 2017~\cite{Amare:2018sxx}. ANAIS-112
consists of 112.5\,kg of NaI(Tl), distributed among nine modules of 12.5\,kg each, which were built by Alpha Spectra starting in 2012.  
The ANAIS-112 crystals offer remarkable light collection that is higher and more homogeneous than that of the DAMA/LIBRA modules, at the level of 15\,npe\,keV$^{-1}$ \cite{Olivan:2017akd} in all modules. A built-in Mylar window allows for calibration with external energy sources only a few keV above the ROI for the testing of the DAMA/LIBRA result. 
Calibrations with $^{109}$Cd sources every 2 weeks enable the correction of small gain drifts, while energy calibration in the ROI profits from energy depositions at 3.2 and 0.87\,keV from the decay of $^{40}$K and $^{22}$Na crystal contaminants, respectively, tagged by coincidences with high-energy $\gamma$ rays in another detector. In particular, the $^{22}$Na calibration confirms highly efficient triggering below 1 keV.

\begin{marginnote}
    \entry{Electron equivalent energy (E$_{ee}$)}{energy scale calibrated with ER energy deposits; given in units of keV}
\end{marginnote} 

ANAIS-112's background model combines input from several analysis techniques~\cite{Amare:2018ndh}. After 6 years of data taking, the average background in the ROI across all nine detectors is 3.23\,cpd\,kg$^{-1}$\,keV$^{-1}$ (Figure~\ref{fig:bkg}). The ANAIS-112 background below 2\,keV exceeds the estimates from the background model.
These events could have a PMT-origin and leak from the filtering protocol, but an unaccounted-for background source is also under consideration, research in both directions is ongoing.
The ANAIS-112 analysis focused on model-independent testing of the DAMA/LIBRA annual modulation. Preliminary annual modulation analyses were released with 1.5, 2 and 3~years of data~\cite{Amare:2019jul,Amare:2019ncj,Amare:2021yyu}; the latest publication included 6 years of data and 625.75\,$\text{kg}\cdot\text{yr}$ of exposure~\cite{Amare:2025dfq}. ANAIS-112 performed the annual modulation analysis in the same energy regions as DAMA/LIBRA in electron equivalent energy ($E_{ee}$), 1--6\,keV and 2--6\,keV. With 6 years of exposure, ANAIS-112 found no evidence of annual modulation, measuring $-0.4\pm2.5$\,cpd\,ton$^{-1}$\,keV$^{-1}$ and $1.1\pm2.5$\,cpd\,ton$^{-1}$\,keV$^{-1}$ in the 1--6 and 2--6\,keV energy regions, respectively.  

In comparing its result to DAMA/LIBRA's, and assuming that the observed signal corresponds to NR, the ANAIS-112 Collaboration accounted for the possibility that the QFs of the detectors used by the two collaborations differ. The DAMA Collaboration~\cite{Bernabei:1996vj} published values of $\mathrm{QF_{Na}}=0.3$ and $\mathrm{QF_I}=0.09$ for sodium and iodine, respectively. According to dedicated measurements of the QF of ANAIS-112 crystals~\cite{Cintas:2024pdu}, systematics associated with the calibration procedure were found to strongly affect the results. Under the assumption of a constant QF, values of $\mathrm{QF_{Na}}=0.210\pm0.003$ and $\mathrm{QF_I}=0.066\pm0.022$ were obtained for ANAIS-112 crystals~\cite{Cintas:2024pdu}. Because the QFs for both nuclei differ by a similar factor of $\sim$3/2 between the two measurements, 
\mbox{2--6\,keV} in DAMA/LIBRA's data would correspond to \mbox{6.7--20\,keV$_\text{nr}$} for sodium and \mbox{22.2--66.7\,keV$_\text{nr}$} for iodine, which in turn would correspond to \mbox{1.3--4.0\,keV} for ANAIS-112. The modulation amplitude, $S_m$, obtained for \mbox{ANAIS-112} in \mbox{6.7--20\,keV$_\text{nr}$} for sodium NR is $0.0\pm2.3$\,cpd\,ton$^{-1}$ per 3.3keV$_\text{nr}$, a value that is incompatible with the DAMA/LIBRA signal at 4.2\,$\sigma$. ANAIS-112 cannot fully explore the region \mbox{1--6\,keV}. However, the analysis of $S_m(E)$ in keV bins was strongly incompatible with DAMA/LIBRA in both the NR and ER energy scales, even when different QF modelings were taken into account. ANAIS-112 took data until January 2026, by which time it had a sensitivity of $\sim5\sigma$. 

\section{INTERPRETING AND TESTING DAMA/LIBRA'S ANNUAL MODULATION}
\label{sec:interpret}

The presence of an annual modulation in DAMA/LIBRA's data is unequivocal. This signal has persisted for more than 25 years, demonstrating a strong statistical significance of 13.7$\sigma$. However, despite substantial efforts to reconcile the signal with the null results from the other experiments, it remains unverified and inconsistent in most of the scenarios considered to date.
We begin this section by summarizing the possible interpretations of the DAMA/LIBRA signal, considering both DM scenarios and potential non-DM systematics. Next, we review the status of the model-independent tests conducted so far, with particular attention to the recently released combined results from COSINE-100 and ANAIS-112. We then discuss the experimental details relevant to this comparison, focusing on one of the key challenges, the QF. 

\subsection{Model-dependent tests}\label{sec:modelDependent}
The modulation amplitude as a function of energy [$S_m(E)$] reported by DAMA/LIBRA-Phase~1 was obtained through a maximum likelihood analysis of the single-hit counts observed for each 1\,keV energy bin for each detector and time interval, assuming Poisson statistics, without background subtraction (unlike the residuals analysis used to present results typically in the 1--6 and 2--6\, keV regions). $S_m(E)$ can be well-described within the canonical WIMP framework under the assumption of purely SI interactions with equal couplings to protons and neutrons. Two regions in the WIMP parameter space provide a good fit to the data~\cite{Savage:2008er}: a light WIMP solution, corresponding to scattering mainly on sodium nuclei with $m_{W} \simeq 12~\mathrm{GeV}$ and $\sigma_{\mathrm{SI}} \simeq 2\times10^{-40}~\mathrm{cm}^{2}$, and a heavier WIMP solution, arising from scattering predominantly on iodine with $m_{W} \simeq 80~\mathrm{GeV}$ and $\sigma_{\mathrm{SI}} \simeq 2.2\times10^{-41}~\mathrm{cm}^{2}$. In the case of SD-interacting WIMPs and the assumption of different coupling to protons and neutrons, the equivalent light- and heavy-mass regions are centered at 
$
(\simeq 10~\mathrm{GeV},\ \simeq 10^{-36}~\mathrm{cm}^{2})$ and
$(\simeq 50~\mathrm{GeV},\  \simeq 8\times10^{-37}~\mathrm{cm}^{2})$, for $\sigma_{\mathrm{SD\!-\!p}}
$ and $(\simeq 10~\mathrm{GeV},\  \simeq 10^{-34}~\mathrm{cm}^{2})$ and 
$(\simeq 50~\mathrm{GeV},\ \simeq 10^{-35}~\mathrm{cm}^{2})$ for $\sigma_{\mathrm{SD\!-\!n}}$.

Baum et al.~\citep{Baum:2018ekm} performed an interesting analysis of the DAMA/LIBRA-Phase~2 result. The analysis energy threshold for DAMA/NaI and DAMA/LIBRA-Phase~1 was 2\,keV, with the modulation signal present in 2--6\,keV. For DAMA/LIBRA-Phase~2, however, data were available dow 1\,keV. In the canonical isospin-conserving SI WIMP-nucleon coupling, and within the standard halo model, we expect the signal to sharply increase below 2\,keV if the observed signal is due primarily to WIMPs with a mass of $\sim$10\,GeV, whereas for a mass of $\sim$100\,GeV the signal is expected to decrease below 2\,keV. DAMA/LIBRA-Phase~2 observed neither scenario but rather a signal that rises smoothly with decreasing recoil energies. Thus, the observed annual modulation signal is not consistent with SI WIMP-nucleon coupling within the standard halo model. 
The DAMA/LIBRA Collaboration~\cite{Bernabei:2019ajy} has argued that its annual modulation signal can still be accommodated within a WIMP framework when relevant astrophysical and detector-related uncertainties (e.g., those on dark halo parameters, QFs, channeling, and nuclear form factors) are properly taken into account. Additionally, when effective field theory operators are introduced, the broader parameter space offers new compatibility scenarios. However, fine-tuning is required to accommodate all experimental results~\cite{Fitzpatrick:2012ix, Anand:2013yka}.

A similar situation arises for SD interactions. The null results from fluorine-based experiments such as PICASSO~\cite{PICASSO:2012ngj}, SIMPLE~\cite{Fernandes:2015fpa}, COUPP~\cite{COUPP:2012jrk}, and PICO~\cite{PICO:2019vsc} have rule out WIMP–proton coupling scenario for DAMA/LIBRA's signal, while xenon-based experiments have rule out WIMP–neutron couplings. Taken together, the combined limits from SI and SD searches indicate that the DAMA/LIBRA modulation cannot be accommodated within standard WIMP models, regardless of whether the coupling is dominated by protons or neutrons. 

No other DM model can reconcile DAMA/LIBRA's signal with the stringent limits set by other experiments~\citep{FreeseColloquium, Green:2011bv}, despite attempts to take out astrophysical dependencies~\cite{Fox:2010bu, Fox:2010bz, Herrero-Garcia:2012arz, Herrero-Garcia:2011dyc}. The DAMA/LIBRA Collaboration has suggested that DM model dependence or subtle detector effects such as QFs~\citep{Bernabei:1996vj} or channeling~\citep{Bernabei:2007hw} are the source of discrepancies between their results and others' null results. Subsequent measurements of QFs and channeling confirmed that these effects cannot account for the discrepancies~\citep{Joo:2018hom, Amare:2025dfq}; for further discussion of QFs, see Section~\ref{sec:QF} .

\subsection{Non-Dark Matter Signal Interpretations}
\label{sec:dama_conundrum}
The DAMA/LIBRA collaboration claims that its result is evidence of DM detection because the observed modulation has a period and phase consistent with DM distributed in the standard halo model, appears only in single-hit events and below 6\,keV, and is distributed consistently among the 25 detectors. The signal cannot be explained by any known background or systematic effect, such as a change in radon levels, temperature, electronics noise, energy calibration, efficiencies, or radioactive backgrounds
~\citep[(for summaries, see~][]{Bernabei:2021kdo, Bernabei:2020mon}). 
Other effects --such as muons and muon-induced events like those induced by phosphorescence in NaI(Tl)~\citep{Nygren:2011xu, DM-Ice:2015aij}; helium leaks into PMTs~\citep{Ferenc:2019esv}; or excesses in potassium~\citep{Ralston:2010bd}, $^{37}$Ar~\citep{McKinsey:2018xdb}, or neutrons and neutrinos~\citep{Davis:2014cja}-- cannot or are unlikely to reproduce a consistent annual modulation signal with the correct magnitude and phase year after year. 

The DAMA/LIBRA Collaboration reports that it has investigated every systematic effect proposed to explain the observed modulation. However, since the data have not been made publicly available, an independent reproduction of the analysis is not currently possible. Experimental aspects related to event selection procedures, efficiencies, energy calibration, time-dependent backgrounds or noises, scintillation QFs, and so forth could be key to solving this puzzle (see Section~\ref{sec:systematics}). 

In contrast, the analysis strategy followed by DAMA/LIBRA is based on study of the residual rates obtained by annually subtracting the average rate.
This procedure is known to be strongly affected by systematic effects in the presence of time-dependent backgrounds~\cite{Buttazzo:2020bto,Messina:2020pnt}, which bias the extracted modulation amplitude and phase. Since DAMA/LIBRA data taking started in the month of December, close to the expected minimum of the DM annual modulation, reproducing a maximum in summer within this analysis approach requires a background component that increases monotonically with time. In contrast, decreasing backgrounds are expected from the decay of radioactive isotopes. COSINE-100 data, analyzed using a similar strategy, yielded a modulation with an amplitude comparable to DAMA/LIBRA's but an opposite phase~\cite{COSINE-100:2022dvc}, as anticipated in the presence of a decreasing background contribution. Although this result remains intriguing, interpreting it as an explanation for the DAMA/LIBRA signal is difficult, as a steadily increasing background in the ROI is challenging to justify on physical grounds.

\subsection{Model-Independent Tests: Annual Modulation Searches with NaI(Tl)}
\label{sec:combinedCOSINEANAIS}
As Carl Sagan noted, extraordinary claims require extraordinary evidence. Detection of DM is an extraordinary claim that necessitates independent verification. For any discovery claim, the scientific method requires that it be replicated under controlled conditions. In the case of the DAMA/LIBRA signal, to date no other experiment has reported a compatible observation. 
Comparisons with experiments using different target media and techniques inherently depend on astrophysical and DM interaction models and cannot be compared on model-independent grounds to discard their result. To eliminate model dependency, the experiment should be recreated with the same target material, NaI(Tl), and the same technique, looking for annual modulations. If the same modulation is not observed in NaI(Tl), then a DM interpretation of the DAMA/LIBRA signal would be ruled out. If no modulation or a different modulation is present, then further investigation would be needed to confirm the signal's origin.

Two experiments, COSINE-100 and ANAIS-112, were designed, constructed, and operated with the explicit goal of directly testing the DAMA/LIBRA claim~(see Sections~\ref{sec:COSINE100} and \ref{sec:ANAIS112}). They used NaI(Tl) as the target medium and carried out annual modulation searches with analysis techniques similar to DAMA/LIBRA's. They also implemented additional improvements in hardware, software, and analysis procedures with the goal of elucidating potential systematics specific to NaI(Tl)-based annual modulation searches~\cite{COSINE-100:2021zqh,Amare:2021yyu,Amare:2025dfq,COSINE-100:2024nfa}. With 6 years of data from each experiment, neither observed an annual modulation. COSINE-100 excluded DAMA/LIBRA's modulations at 3$\sigma$ and ANAIS-112 at 4$\sigma$. Although the two experiments have similar sensitivities, COSINE-100 observed a small upward fluctuation early in its data, making the final result statistically less significant.

The COSINE-100 and ANAIS-112 Collaborations also performed a combined analysis of their data, using detailed detector simulations and taking into account both known time-dependent and time-independent backgrounds for each experiment~\cite{ANAIS-112:2025fne}. 
A detailed analysis was first carried out with 3 years of data from both experiments to benchmark potential systematics among the analysis procedures or differences between the two experiments. No statistically significant differences were among the results obtained from the different analysis procedures, namely frequentist ($\chi^2$~minimization of least-squares, used by ANAIS-112) and Bayesian approaches [Markov chain Monte Carlo (MC), used by COSINE-100], fitting of the event rate before versus after subtracting the time-dependent backgrounds from known long-lived isotopes, and simple combination of the results from the two experiments. 

\begin{figure}
    \centering
    \includegraphics[width=1\textwidth]{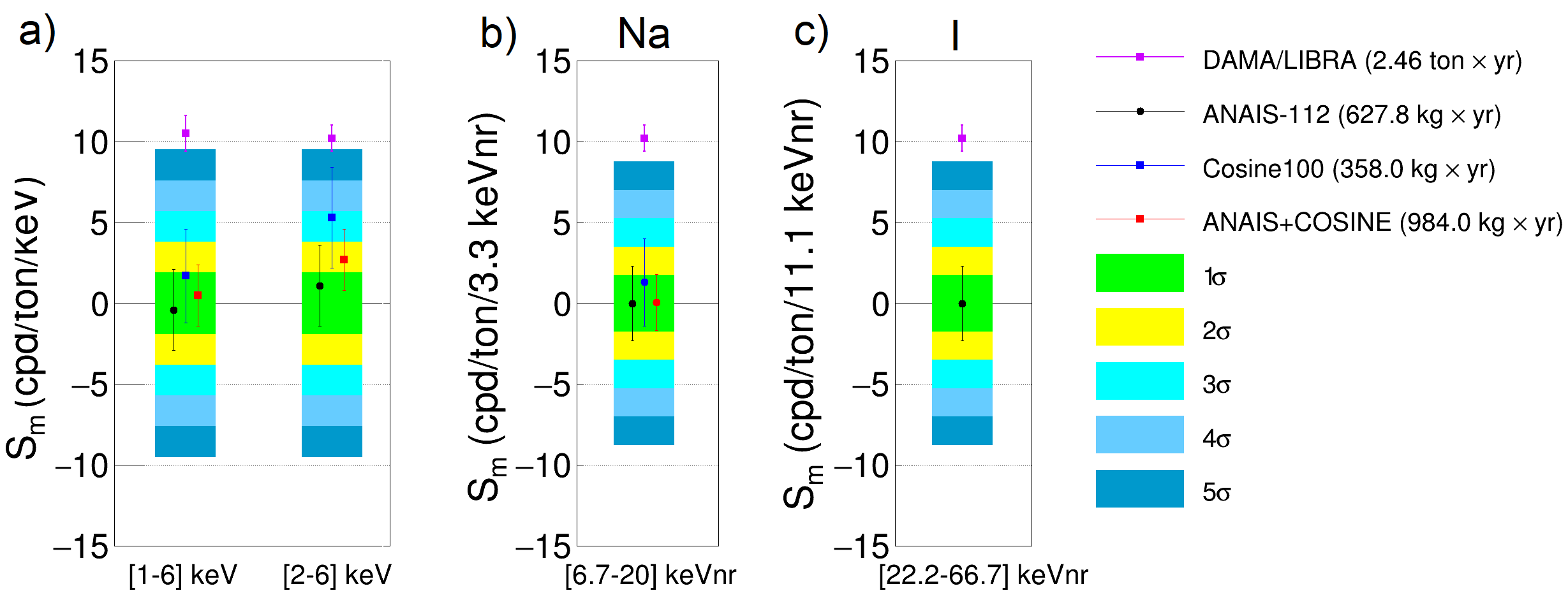}
    \caption{Combined modulation result~\cite{ANAIS-112:2025fne} of the full COSINE-100 data set (6.4 years)~\cite{COSINE-100:2024nfa} and the ANAIS-112 data set (6 year)~\cite{Amare:2025dfq}. Independent results from both experiments and DAMA/LIBRA are shown in {\it (a)} the electron equivalent energy regions and {\it (b), (c)} the NaI nuclear recoil scales. For the latter, constant QF$_\text{Na}=0.2$ {\it (b)} and QF$_\text{I}=0.06$ {\it (c)} for ANAIS crystals were considered~\cite{Cintas:2024pdu}. Abbreviation: QF, quenching factor.}
     \label{fig:combined}
\end{figure}

In the combined analysis for the 3 year data set of COSINE-100 and ANAIS-112, the best-fit values obtained for the modulation amplitude are $-0.2\pm2.6$\,cpd\,ton$^{-1}$\,keV$^{-1}$ in the 1--6\,keV energy range and $2.1\pm2.8$\,cpd\,ton$^{-1}$\,keV$^{-1}$ in the 2--6\,keV range. These results are consistent with no modulation and incompatible with DAMA/LIBRA's modulation amplitude at significance levels of 3.7$\sigma$ for 1--6\,keV and 2.6$\sigma$ for 2--6\,keV.  
For the 6 year combined data (for a total exposure of 984\,kg$\cdot$yr), for which only the simple combination was performed after it was demonstrated to be equivalent to the detailed combination, the best-fit values for the modulation amplitude are $0.5\pm1.9$\,cpd\,ton$^{-1}$\,keV$^{-1}$ in the 1--6\,keV range and $2.7\pm2.1$\,cpd\,ton$^{-1}$\,keV$^{-1}$ in 2--6\,keV. These results are consistent with no modulation and incompatible with DAMA/LIBRA's modulation amplitude at significance levels of 4.7$\sigma$ and 3.5$\sigma$. Figure~\ref{fig:combined} summarizes the current status of the DAMA/LIBRA testing by COSINE-100 (6.4 year full data set), ANAIS-112 (6 year data), and the two experiments combined. 

\subsection{Caveats for Model Independence}
\label{sec:systematics}
If detectors were ideal devices, the DM signal would be the same in the same target. In actual detectors, however, several factors could affect the measured signal. We use the total number of photons detected as a proxy for the total energy deposited in the NaI(Tl) detectors (signal). Then, the conversion of the deposited energy to a detected signal under different conditions must be characterized and understood. In this section, we summarize detector systematics that could affect the comparison between the DAMA/LIBRA, ANAIS and COSINE-100 experiments. 

\subsubsection{Detector response linearity}
NaI(Tl) scintillators are well-known to be nonproportional, so proper energy calibration is crucial. The light yield of NaI(Tl) changes at the level of a few percent up to 20\,keV~\cite{PhysRev.122.815, IEEETNS1997, Moses:2001vx, IEEETNS2008, IEEETNS2009, IEEETNS2009b, Khodyuk:2010ydw, Payne:2011zz, COSINE-100:2021mrq}, and it is difficult to model.
Although they follow different calibration protocols, the DAMA/LIBRA, ANAIS-112, and COSINE-100 experiments all use the 3.2\,keV $^{40}$K line to anchor the energy scale in the ROI for the annual modulation signal.

\subsubsection{Event selection efficiencies}
Event selection efficiencies are a crucial experimental parameter, as the rates in the ROI are dominated by anomalous scintillation. The three experiments use different event selection strategies, and their selection efficiencies as a function of energy behave quite differently. ANAIS-112 and COSINE-100 use machine learning filtering, while DAMA/LIBRA uses two pulse-shape parameters. Moreover, different populations of anomalous events seem to be present in each experiment. The efficiencies are determined from ER calibration events for the COSINE-100 and DAMA/LIBRA experiments, but ANAIS-112 uses NR and ER events~\cite{Amare:2025dfq}. 

\subsubsection{Crystal contamination levels}
The contamination levels of DAMA/LIBRA's crystals are below 620\,$\mu$Bq\,kg$^{-1}$ for $^{40}$K and 5--30\,$\mu$Bq\,kg$^{-1}$ for $^{210}$Pb, significantly lower than those of COSINE-100 and ANAIS-112, which are 700--1,330\,$\mu$Bq\,kg$^{-1}$ for $^{40}$K and 700--3,150\,$\mu$Bq\,kg$^{-1}$ for $^{210}$Pb. As a result, COSINE-100 and ANAIS-112 have a larger background in the ROI (Figure~\ref{fig:bkg}). 
COSINE-100 benefits from a lower contribution from $^{40}$K in the ROI because the active veto from the surrounding liquid scintillator removes these events. However, the background above 4\,keV is larger than for ANAIS-112. 
Another relevant contribution to the background in the ROI is tritium, produced by cosmic neutrons before the detectors were installed underground. Tritium was identified within the ANAIS and COSINE crystals at a level of 0.09--0.20\,mBq\,kg$^{-1}$~\cite{Amare:2018ndh,COSINE-100:2019rvp}, consistent with estimated production rates~\cite{Amare:2017roa,COSINE-100:2019rvp}.
Despite their larger background, ANAIS-112 and COSINE-100 are sensitive to the DAMA/LIBRA signal. The two collaborations have thoroughly checked this sensitivity by developing dedicated toy MC simulations of equivalent experiments, injecting the DM signal, and recovering the unbiased modulation amplitude with the expected uncertainty. Within these MC estimates, the data analysis incorporated the event selection efficiency uncertainties as a possible systematic effect, showing a negligible contribution unless the efficiencies are modulating with exactly the opposite phase, which is indeed difficult to justify~\cite{Amare:2025dfq}. 

\begin{marginnote}
    \entry{NR energy (E$_{nr}$)}{energy scale calibrated with NR energy deposits; given in units of keV$_{nr}$}
\end{marginnote} 

\subsubsection{Quenching factors}
\label{sec:QF}
The largest systematic effect is likely the scintillation QF in NaI(Tl). 
Only a fraction of the energy deposited by a particle interacting in a NaI(Tl) crystal results in the emission of scintillation light. This fraction differs for an electron, an $\alpha$ particle, or a heavier nucleus, and the conversion between the light collected and energy deposited (i.e.,~the calibration of the energy scale) also differs. 
Typically, $\gamma$ rays, which produce ER events, are used for calibration; thus, the corresponding energy scale should be referred to as electron-equivalent energy (E\textsubscript{ee}), whereas NR energy (E\textsubscript{nr}) should require a different calibration.

In 1996, the DAMA Collaboration~\cite{Bernabei:1996vj} performed an early measurement of the sodium and iodine scintillation QFs in NaI(Tl) detectors, following Fushimi et al.~\cite{Fushimi:1993nq}. It measured the response of a NaI(Tl) detector to neutrons emitted by $^{252}$Cf and fitted to a spectral shape (MC justified), under the assumption of constant QFs.  Using this approach, DAMA found  QF$\mathrm{_{Na}}=0.30\pm0.01$ at 6.5--97\,keV\textsubscript{nr} and QF$\mathrm{_{I}}=0.09\pm0.01$ at 22--330\,keV\textsubscript{nr}, significantly differed from the Fushimi et al. values of QF$\mathrm{_{Na}}=0.40\pm0.2$ and QF$\mathrm{_{I}}=0.05\pm0.02$.
ANAIS is following a similar approach, comparing the measurements with a full Geant4 simulation of neutron transport through the experimental setup and considering various QF models~\cite{Yanguas:2025vny}.

Updated measurements have been performed through the use of a monoenergetic neutron beam scattering off the detector~\cite{Collar:2013gu, Xu:2015wha, Lee:2024unz, Cintas:2024pdu}. These include measurements of Alpha Spectra crystals similar to those used by ANAIS and COSINE. 
Detection of the scattered neutron at a certain angle allows the NR energies produced in the NaI(Tl) to be selected, which in turn enables investigation of the energy dependence of the QF.
The measurements take  following this approach have consistently produced QF$\mathrm{_{Na}}$ and QF$\mathrm{_{I}}$ values that are lower than those reported by DAMA, with QF$\mathrm{_{Na}}$ decreasing with decreasing recoil energy. The nonproportionality of the light yield must be carefully corrected for these measurements~\cite{Cintas:2024pdu}. 
According to a preliminary analysis of the QF of NaI(Tl) crystals with differing thallium content, no large impact on the QF value has been observed~\cite{Bharadwaj:2023aoz}. 
Note that most of the measurements performed with monoenergetic neutron beams yield lower QF values than those reported by DAMA. Additionally, ANAIS's measurements using neutrons emitted by $^{252}$Cf, which were analyzed by fitting the spectral shape to a MC-produced template, also resulted in lower QF values, consistent with those obtained using a monoenergetic neutron beam~\cite{Yanguas:2025vny}.

Both, ANAIS-112 and COSINE-100 have ruled out the DAMA/LIBRA signal with their modulation searches, even when QF differences were taken into account (Figure~\ref{fig:combined}). Although more work remains to be done to gain a better understanding of the origin of the discrepancies in QF measurements, the QF cannot explain the fact that DAMA/LIBRA observed an annual modulation but other experiments (including ANAIS-112 and COSINE-100) found no signal in the 1--6\,keV range.

\subsubsection{Other experimental differences}
Although less important than those described above, several other relevant experimental differences among the three experiments exist. We discuss each below.

\textbf{Muon veto.} Unlike the DAMA/LIBRA experiment, both COSINE-100 and ANAIS-112 have muon-tagging systems, installed in laboratories with a lower rock overburden than LNGS. The residual underground muon flux has a seasonal modulation~\cite{MACRO:1997teb,MINOS:2009njg,Adamson:2014xga,DoubleChooz:2016sdt,GERDA:2016lhn,DayaBay:2017lpf,Borexino:2018pev,OPERA:2018jif,LVD:2019zlh,COSINE-100:2020jml}, and muons were proposed as a possible origin of the DAMA/LIBRA signal, although the collaboration has always disregarded that idea~\cite{Bernabei:2012wp}. 

\textbf{Data acquisition, waveform sampling, and triggering.}
The three experiments follow different data acquisition strategies. ANAIS-112 uses a higher sampling rate and a shorter time window than COSINE-100 and a larger coincidence window for the module trigger between the two PMTs signals than DAMA/LIBRA, resulting in greater trigger efficiency. The sampling rate, ADC-bit resolution, integration window, and trigger efficiency differ between the three experiments. ANAIS-112 installed a new data acquisition system (ANOD) in 2024 to evaluate some of these effects, namely a longer digitization window and dead-time effects. Data analysis is still ongoing. ANOD data are expected to provide relevant input that will enable an improved rejection of anomalous scintillation events~\cite{Yanguas:2025vny}. The implications of these differences should be analyzed, but it is unlikely that they could explain either the nonobservation of modulation by the COSINE and ANAIS experiments or the modulation observed by DAMA/LIBRA. 

\subsection{Data Sharing and Cross Checks}
The ANAIS-112 and COSINE-100 experiments openly share their processed data to allow others to reconstruct their fitting and published results. The data can be found on the ORIGINS Excellence Cluster website
(\url{https://www.origins-cluster.de/odsl/dark-matter-data-center}).
Such efforts illustrate the path toward a more transparent and robust framework for future DM searches. However, to advance more general open science-driven scenarios, it will be necessary to establish a coordinated and widely supported strategy.

\section{FUTURE OF SODIUM IODIDE DARK MATTER SEARCHES}
\label{sec:projects}
\begin{figure}
    \centering
    \begin{minipage}{0.48\textwidth}
        \centering
        \includegraphics[width=\textwidth]{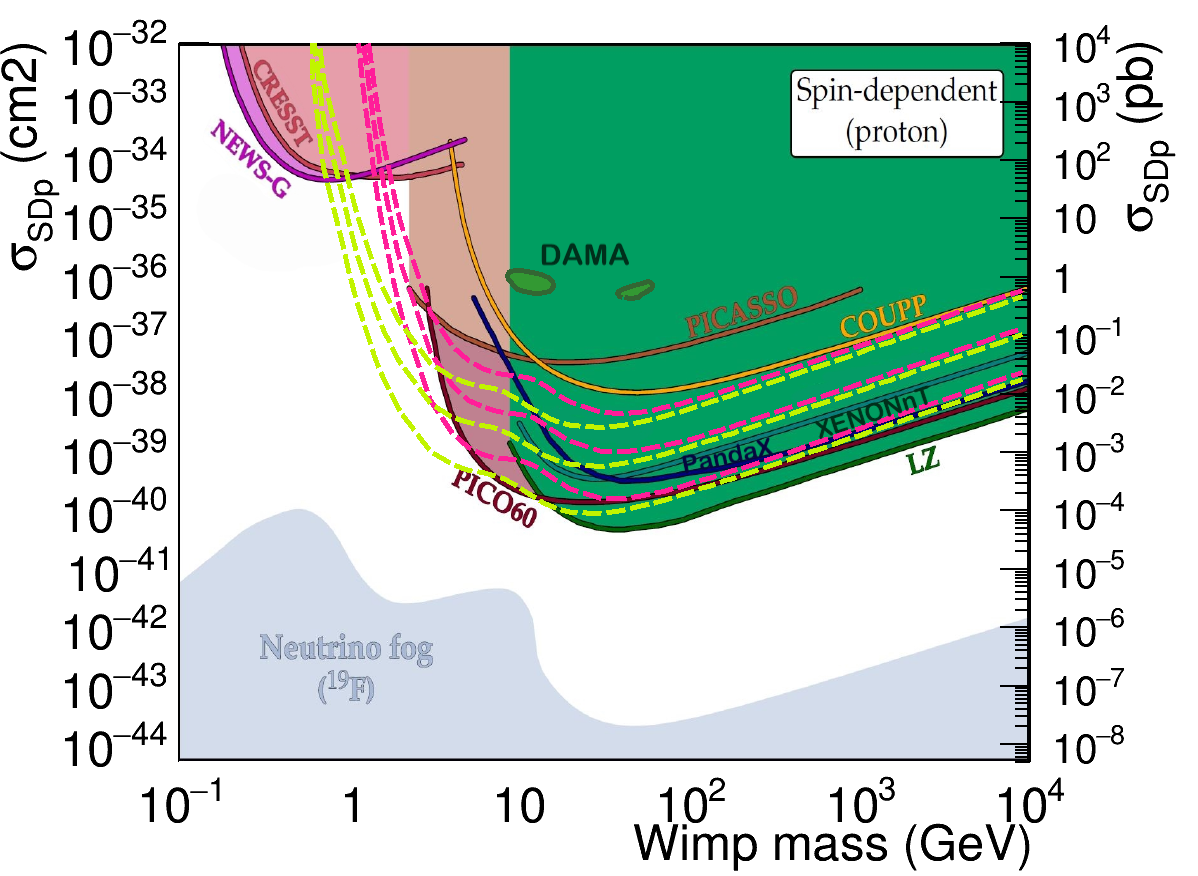}
    \end{minipage}
    \hfill
    \begin{minipage}{0.48\textwidth}
        \centering
        \includegraphics[width=\textwidth]{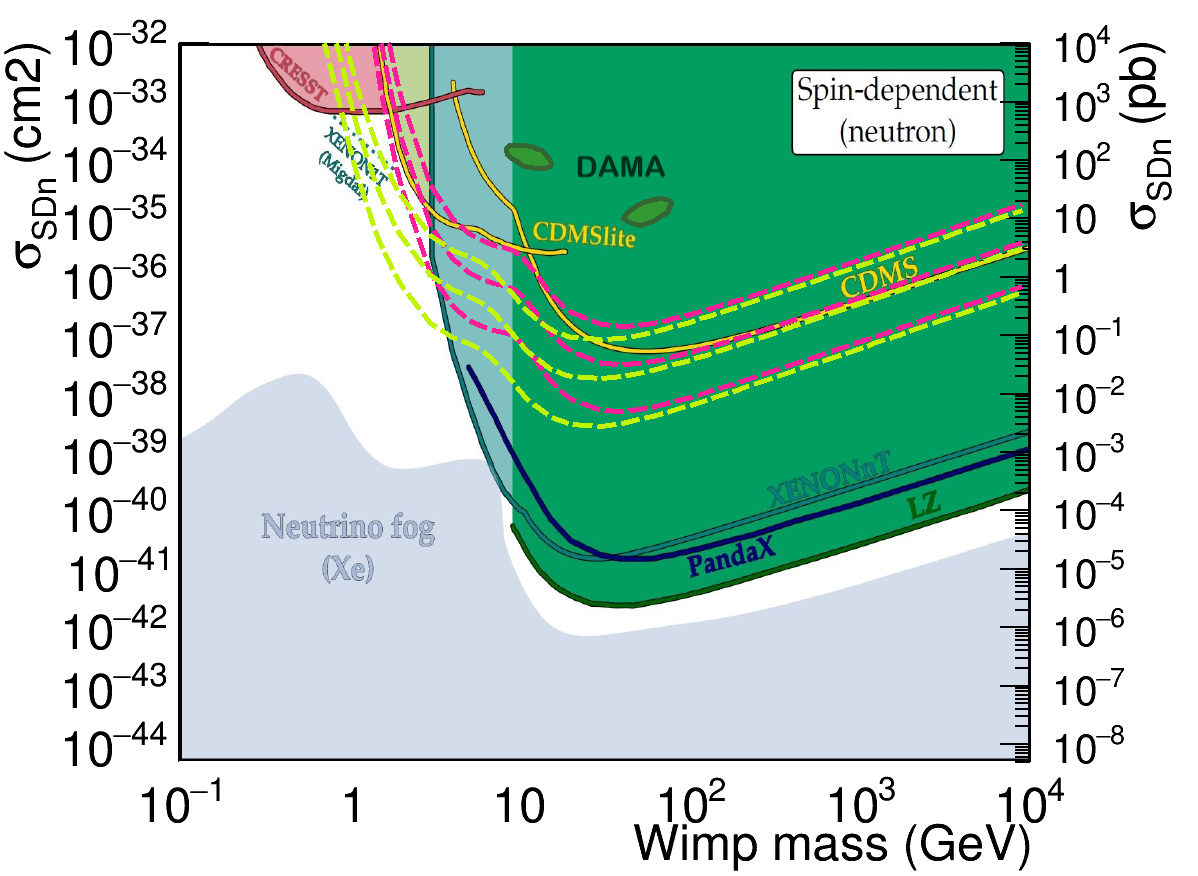}
    \end{minipage}
    \caption{Experimental reach for spin-dependent WIMP–proton (left) and WIMP–neutron (right) interactions. The dashed pink and green curves represent the projected sensitivities for a next-generation NaI experiment operated with energy thresholds of 250\,eV and 150\,eV, respectively, assuming a background level of 0.2\,cpd\,kg$^{-1}$\,keV$^{-1}$. In each panel, the three curves correspond to exposures of 1\,kg$\cdot$yr, 10\,kg$\cdot$yr, and 200\,kg$\cdot$yr. 
    The neutrino fog is shown for $^{19}$F (left) and xenon (right) as reference. The colored regions indicate parameter space excluded by the corresponding experiments. Abbreviation: WIMP, weakly interacting massive particle. Figure adapted with permission from \url{https://github.com/cajohare/DirectDetectionPlots.}}
     \label{fig:ANAIS+}
     \label{fig:lowMassWIMPSensitivity}
\end{figure}

NaI detectors remain competitive as DM detectors for all the same reasons they became competitive decades ago. With no DM detected, the parameter space has significantly broadened. Large experiments are setting aside ER-versus-NR discrimination and significantly reducing the energy threshold to look for low-mass WIMPs. In this arena, NaI detectors are sensitive to SD WIMP-proton and WIMP-neutron interactions and could provide competitive limits with modest improvements in threshold and background (Figure~\ref{fig:ANAIS+}). Next-generation experiments using NaI(Tl) will follow different experimental avenues to increase the sensitivity for DM searches, as summarized in this section. 

\subsection{Next-Generation Experiments Using NaI(Tl) with Photomultiplier Tube Readout}

Several experiments that are currently underway or in an advanced development stage continue to
use PMTs to read out NaI(Tl) detectors. These build on the successes of the previous generation of
experiments such as COSINE-100 and ANAIS-112 while pursuing improved sensitivity through
a range of strategies, including deeper or multiple underground sites, lower-background crystals,
increased detector mass, and refined detector geometries.

\subsubsection{COSINE-100U}
\label{sec:COSINE100U}
The COSINE-100U experiment consists of eight NaI(Tl) crystals with a total mass of 99.1\,kg, and is located at Yemilab, an underground laboratory that opened in 2022~\cite{Park:2024sio, Kim:2024xyd}. The detectors used in COSINE-100 were resurfaced and tapered at the ends to 3 inches to match the diameter of the PMTs. The encapsulation was redesigned to eliminate the quartz light guides. These improvements resulted in a light yield of up to 22\,npe\,keV$^{-1}$, approximately 50\% higher than the 15\,npe\,keV$^{-1}$ light yield observed in COSINE-100~\cite{Choi:2020qcj, NEON:2024bsc}. The crystal resurfacing and reencapsulation reduced the number of $\alpha$ particle-emitting surface contamination from $\sim$90\,nBq\,cm$^{-2}$ observed in COSINE-100 to $\sim$20\,nB\,cm$^{-2}$~\cite{COSINE-100:2024ola}. 

COSINE-100U began taking physics data in September 2025~\cite{Lee:2024wzd}. It operated at room temperature for approximately 6 months, after which it was cooled down to operate at \mbox{$-30$\degree C} so as to enhance light yield and improve PSD for NR events. Earlier measurements indicate an increase in light yield of $\sim$5\% compared with operation at room temperature for ER events, and an $\sim$9\% increase in the QF for $\alpha$ particles was observed, suggesting potential further improvement in light yield for NR events~\cite{Lee:2021aoi}. These upgrades willsignificantly improve the experiment's ability to probe low-mass DM candidates~\cite{Lee:2024wzd}.

\subsubsection{SABRE}
\label{sec:SABRE}
The SABRE (Sodium Iodide with Active Background REjection) Collaboration plans to install two experiments, one at LNGS in the Northern Hemisphere and another in the Southern Hemisphere at the Stawell Gold Mine in Victoria, Australia. The DM signal should be same in both hemispheres, whereas site-dependent effects such as those caused by seasonal variation are expected to be different.
Having one site at LNGS, where DAMA/LIBRA was located, may offer a clue as to what DAMA/LIBRA may have been seeing. 
SABRE's goal is to develop beyond-state-of-the-art purification of raw NaI material by using a zone-refinement process, which can significantly reduce impurities. The crystals would then be grown using the Bridgman method by Radiation Monitoring Devices (Boston, Massachusetts, USA). To date, 
SABRE has achieved a very low $^{nat}$K level of $4.3\pm0.2$\,ppb but the $^{210}$Pb content in the crystal bulk is still high ($340\pm40$\,$\mu$Bq\,kg$^{-1}$ and $461\pm5$\,$\mu$Bq\,kg$^{-1}$ for two different crystals)~\cite{Antonello:2020xhj}.

\label{sec:SABRENorth}
\textbf{SABRE-North.} The SABRE-North experimental plan consists of an array of nine crystals, each with a mass of $\sim$5\,kg. In the proof-of-principle phase at LNGS, SABRE-North used an active veto (liquid scintillator) to characterize the detector performance, but an analysis showed that, with sufficiently clean crystals and shielding, the veto might be unnecessary for target background levels of 0.3--0.5\,cpd\,kg$^{-1}$\,keV$^{-1}$ \cite{Calaprice:2022vte}. Characterization of the NaI-33 crystal in the proof-of-principle setup yielded values of $1.20\pm0.05$\,cpd\,kg$^{-1}$\,keV$^{-1}$ in the 1--6\,keV range and $1.39\pm0.02$\,cpd\,kg$^{-1}$\,keV$^{-1}$ in the setup without an active veto~\cite{Antonello:2020xhj}. The reported light collection in the proof-of-principle phase was 11\,npe\,keV$^{-1}$, although the SABRE-North Collaboration expects to increase this value by using higher-efficiency PMTs.

\label{sec:SABRESouth}
\textbf{SABRE-South.} SABRE-South will be located in the Stawell Underground Physics Laboratory, located in the Stawell Gold Mine under a rock overburden of 2,900\,m.w.e. The 
SABRE-South crystals will be installed within a liquid scintillator veto and plastic scintillators for muon tagging~\cite{SABRESouth:2024bpv}. Two designs are under consideration: one with seven crystals of $\sim$7\,kg each, for a total mass of $\sim$50\,kg, and another with seven crystals of $\sim$5\,kg each, for a total mass of $\sim$35\,kg. The former set of crystals would be grown in the Bridgman method by the Shanghai Institute of Ceramics of the Chinese Academy of Sciences, while the latter would be grown by Radiation Monitoring Devices after zone refinement of the raw powder (following the same procedure as SABRE-North). The expected background rate in the 1--6\,keV region, after veto cuts are applied, is estimated to be $\sim$0.72\,cpd\,kg$^{-1}$\,keV$^{-1}$ \cite{SABRESouth:2024bpv}.

\subsubsection{PICOLON}
\label{sec:PICOLON}
The PICOLON [Pure Inorganic Crystal Observatory for LOw-energy Neutr(al)ino] DM search project has successfully grown crystals with very low $^{210}$Pb contamination (below 6\,$\mu$Bq\,kg$^{-1}$) and $^{nat}$K below 20\,ppb by using a hybrid purification method combining recrystallization and ion exchange resins. PICOLON plans to build a 250\,kg NaI(Tl) experiment in the KamLAND area of the Kamioka Underground Laboratory at a depth of 2,700\,m.w.e.~\cite{Fushimi:2021mez}.

Recently, PICOLON achieved the same radiopurity as the DAMA/LIBRA detectors. A 6-month measurement with a crystal mass of 1.34\,kg confirmed the long-term stability of the counting rate and enabled the experiment to perform its first annual modulation search~\cite{PICOLON:2025edl}. The light collection achieved is $\sim$10\,npe\,keV$^{-1}$.

\subsubsection{COSINE-200}
\label{sec:COSINE200}
The COSINE Collaboration has made significant progress in producing new ultrapure NaI(Tl) detectors~\cite{COSINE_recrystalize, Shin:2020bdq, Lee:2023jbe, Shin:2023ldy}. The COSINE-200 experiment will grow 200\,kg of highly radiopure NaI(Tl) detectors, assembled in a low-background environment. It has produced 400\,kg of low-background NaI powder~\cite{Shin:2023ldy}, and grown NaI crystals with purity levels of 10\,$\mu$Bq\,kg$^{-1}$ of $^{210}$Pb and 8~ppb of $^{nat}$K~\cite{Lee:2023jbe} --comparable to or below DAMA's crystals. COSINE-200 will employ all that is learned from COSINE-100U, but with a lower expected background rate (Figure~\ref{fig:bkg}).

\subsection{Next-Generation Experiments Using Sodium Iodide Scintillating Bolometers: COSINUS}
\label{sec:COSINUS}

COSINUS (Cryogenic Observatory for SIgnatures seen in Next-generation Underground Searches) uses both heat and light generated from particle interactions in NaI crystals. This hybrid detection scheme allows the experiment to discriminate NR from ER~\cite{COSINUS:2021onk}. 
The heat channel is quite sensitive to NR energy depositions and could reach very low energy thresholds. However, attaining a large exposure could be difficult. 

The COSINUS experiment uses undoped NaI as its target medium. The detectors must be cooled to milli-Kelvin temperatures in order to operate as bolometers. The experiment is currently being commissioned at the LNGS~\cite{COSINUS:2023kqd}. The COSINUS Collaboration has developed a novel ``remoTES'' design to read out the heat channel, using a transition edge sensor (TES) attached to an external wafer substrate that is thermally coupled to an NaI crystal~\cite{COSINUS:2021onk}. A more conventional silicon bolometer is used for the light channel.
The COSINUS setup at LNGS consists of a dry cryostat inside a water tank that acts as a shield and veto for many background components. COSINUS's first detector configuration will consist of eight modules, 30\,g each, which will aim for a threshold of 1~keV.

\begin{marginnote}
    \entry{Transition edge sensor (TES)}{very sensitive thermometer consisting of a superconductor film stabilized in the superconductive-normal transition}
\end{marginnote} 

Among the goals of COSINUS is to perform model-independent testing of the DAMA/LIBRA DM signal by comparing the unquenched NR spectrum corresponding to such a signal (obtained by unfolding the modulation spectrum measured by DAMA/LIBRA, which requires knowledge of the DAMA/LIBRA crystals' QF) with COSINUS's sensitivity. This method relies on the total interaction rate in NR energy, without requiring a modulation analysis, but there are some caveats~\cite{COSINUS:2025xpt}.

\subsection{Next-Generation Experiments Using NaI/NaI(Tl) with Silicon Photomultiplier Readout}
Crystal contamination is the dominant source of background in the current-generation NaI detectors. However, as the radiopurity of the crystals improves, PMTs will become the primary concern. Moreover, PMTs produce spurious events that are difficult to get rid of. Efforts are underway to replace PMTs with silicon photomultipliers (SiPMs).

\begin{marginnote}
    \entry{SiPMs}{silicon photomultipliers.}
\end{marginnote} 

\subsubsection{ANAIS+}
\label{sec:ANAIS+}
The ANAIS+ Collaboration (the next phase of ANAIS-112) is currently investigating the replacement of conventional PMTs with SiPMs operated at cryogenic temperatures of $\sim$100\,K. The use of SiPMs would eliminate the PMT-induced light noise observed by ANAIS-112. The high quantum efficiency would allow for improved light collection, while the light output of undoped NaI at these temperatures is expected to be higher than that of NaI(Tl). These improvements offer the potential to reach an energy threshold as low as $\sim$100\,eV, which will provide a competitive sensitivity to SI and SD interactions for sub-GeV WIMPs (Figure~\ref{fig:ANAIS+}) if combined with highly radiopure crystals. 

The first ANAIS prototypes consist of cubic crystals, measuring 1\,inch per side, of undoped CsI and NaI, with four of the crystal faces coupled to SiPM arrays. Light collection above 22\,npe\,keV$^{-1}$ has been measured for CsI, with effective threshold at the keV level, in the first nonoptimized readout tests. In the case of NaI, very slow scintillation time constants have been observed, making readout difficult. 
Strong scintillation dependencies on temperature are expected for intrinsic inorganic scintillators; therefore, identifying the optimal operating temperature will be crucial for ensuring a high-performance detector.
The first tests underground and inside LAr as the active veto are being planned for the coming months.

\subsubsection{ASTAROTH}

The ASTAROTH project is developing NaI(Tl) scintillators coupled to SiPMs that will be operated at cryogenic temperatures. The first ASTAROTH prototype~\cite{Martinenghi:2025ckb} consists of a cylindrical NaI(Tl) crystal (50\,mm in diameter and height, $\sim$360\,g) with one face coupled to an $8\times8$ SiPM matrix developed by FBK. The detector achieved a gross light yield of 7.2\,npe\,keV$^{-1}$, corresponding to 4.5\,npe\,keV$^{-1}$ after cross-talk correction. 
An energy resolution of $(26.6\pm0.4)\%$~FWHM was measured at 59.5~keV, and the effective trigger threshold was approximately 0.5\,keV.

\section{CONCLUSIONS}
\label{sec:Conclusions}

NaI detectors have played a significant role in particle physics. Their relatively large size, high radiopurity, high light yield, good energy resolution, stability, and ease of operation over long time periods make them ideal DM detectors. They led the field of direct detection DM searches in size and sensitivity in the early days, and they have helped shape the direction of the field. 

The DAMA/LIBRA Collaboration's observation of an unequivocal annual modulation and its claim of DM discovery, while controversial, have stimulated creativity and innovation. The controversy created urgency and context in which to carefully investigate experimental and detector systematics, come up with extensions to DM models, and consider whether the simple SI interactions and standard halo models were worth modifying. At the same time, larger experiments with ER rejection capabilities came online, steadily reducing the parameter space for a viable explanation for DAMA/LIBRA's signal arising from DM. 

Today, we are close to solving the DAMA/LIBRA puzzle. Results from the ANAIS and COSINE experiments strongly disfavor a DM interpretation of the reported annual modulation, although further investigation is needed to clarify its origin. The ANAIS and COSINE Collaborations are continuing their efforts with improved detectors. SABRE, PICOLON, and COSINUS will come online with different strategies to determine what DAMA/LIBRA may have been seeing. Although the DAMA/LIBRA experient is finished, efforts are underway to independently test the detectors that are carefully kept at LNGS.

NaI-based detectors will continue to play a role that goes well beyond addressing the DAMA/LIBRA puzzle. They complement the current best-performing technologies employed in DM searches, namely those based on LXe and LAr; are sensitive to sub-keV recoils; and have the potential to observe light WIMPs with SD coupling. NaI detectors will continue to be competitive for all the same reasons as in the 1990s when we began searching for DM.

The results from DAMA/LIBRA, and the challenges of testing them, highlight the critical importance of reproducibility in scientific methodology. As contemporary DM experiments become increasingly large and complex, replicating results will be more difficult, if not impossible. It is essential that we adopt best practices for sharing experimental details, data, and analytical tools. Implementing coordinated strategies to navigate the path to discovery would greatly benefit the DM research community.

New experiments with the potential to make groundbreaking discoveries are being launched. While DM detection could be just around the corner, a definitive resolution of the DAMA/LIBRA anomaly should provide key lessons on how to transform a potential signal into a validated discovery.

\begin{summary}[SUMMARY POINTS]
\begin{enumerate}
    \item Thallium-doped sodium iodide [NaI(Tl)] detectors were adapted early on as competitive dark matter (DM) detectors. 
    \item The DAMA/NaI experiment made its first claim of detection of annual modulation as consistent with DM in 1997, followed by its successor experiment, DAMA/LIBRA. The detected signal has been ruled out as standard weakly ineracting massive particles (WIMPs) with non-NaI-based experiments.
    \item The NaI(Tl)-based experiments ANAIS-112 and COSINE-100 do not see annual modulation, strongly indicating that DAMA/LIBRA's annual modulation is not caused by DM.
    \item NaI-based detectors remain competitive for their potential to achieve sub-keV energy thresholds, ease of scaling up to high masses, and unique sensitivity to multiple DM models following different experimental strategies.
\end{enumerate}
\end{summary}

\begin{issues}[FUTURE ISSUES]
\begin{enumerate}
    \item The source of the annual modulation observed by DAMA/LIBRA is still unknown. 
    \item Experiments with particle identification capabilities, such as COSINUS or those located at Gran Sasso National Laboratory (LNGS), like COSINUS and SABRE-North, may offer clues as to the origin of DAMA/LIBRA's annual modulation.
    \item The NaI(Tl) detectors used in the DAMA/LIBRA experiment are now in the custody of INFN. The community should take this opportunity to verify the detectors' key performance factors, such as radiopurity, the scintillation quenching factor, energy calibration, and stability against environmental changes at LNGS.
    \item NaI detectors continue to offer a competitive platform for performing searches for a wide variety of DM models and other rare events.
\end{enumerate}
\end{issues}

\section*{DISCLOSURE STATEMENT}
M.M. and M.L.S. are members of the ANAIS and ANAIS+ Collaborations. R.H.M. is a member
of the DM-Ice17 and COSINE-100 Collaborations.

\section*{ACKNOWLEDGMENTS}
M.M and M.L.S acknowledge fiancial support from Ministerio de Ciencia e Innovación/
Agencia Estatal de Investigación (MCIN/AEI/10.13039/501100011033; grant number PID2022-
138357NB-C21), Gobierno de Aragón, and the European Social Fund (Nuclear and Astroparticle
Physics Group), as well as funds from European Union NextGenerationEU/PRTR (Planes com-
plementarios, Programa de Astrofísica y Física de Altas Energías). R.H.M. acknowledges financial
support from the National Science Foundation (award number PHY-1913742).



\bibliography{NaI_DM}
\bibliographystyle{ar-style5}

\end{document}